# Static Non-Reciprocity in Mechanical Metamaterials

Corentin Coulais[1,2,*], Dimitrios Sounas[3], Andrea Alù[3]

[1]AMOLF, Science Park 104, 1098 XG Amsterdam, The Netherlands

[2]Huygens-Kamerlingh Onnes Lab, Universiteit Leiden, PObox 9504, 2300 RA Leiden, The Netherlands

[3]Department of Electrical and Computer Engineering, The University of Texas at Austin, Austin, Texas 78701, USA

[*]Present address: Van der Waals – Zeeman Institute, Institute of Physics, Universiteit van Amsterdam, Science Park 904, 1098 XH Amsterdam, The Netherlands

**Reciprocity is a fundamental principle governing various physical systems, which ensures that the transfer function between any two points in space is identical, regardless of geometrical or material asymmetries. Breaking this transmission symmetry offers enhanced control over signal transport, isolation and source protection[1-6]. So far, devices that break reciprocity have been mostly considered in dynamic systems, for electromagnetic, acoustic and mechanical wave propagation associated with spatio-temporal variations. Here we show that it is possible to strongly break reciprocity in static systems, realizing mechanical metamaterials[7-16] that, by combining large nonlinearities with suitable geometrical asymmetries, and possibly topological features, exhibit vastly different output displacements under excitation from different sides, as well as one-way displacement amplification. In addition to extending non-reciprocity and isolation to statics, our work sheds new light on**

**the understanding of energy propagation in non-linear materials with asymmetric crystalline structures and topological properties, opening avenues for energy absorption, conversion and harvesting, soft robotics, prosthetics and optomechanics.**

By pushing an object on side *A*, we move the other side *B* by a certain amount. Daily experience tells us that if we now push the opposite side *B* with the same force, side *A* moves by the same amount. In other words, if an object transmits motion in one direction, it typically does also in the opposite one. This basic property of static mechanical systems is a direct consequence of Maxwell-Betti's reciprocity theorem, a widely known result at the foundations of mechanical engineering[17-19], with important consequences for the analysis and design of a wide range of mechanical systems[20,21]. Maxwell-Betti's theorem is mathematically formulated as

$$F_A u_{B \to A} = F_B u_{A \to B} \ , \qquad (1)$$

where $F_A$ ($F_B$) is the applied force at point *A* (*B*) and $u_{A \to B}$ ($u_{B \to A}$) the displacement at point *B* (*A*) induced by $F_A$ ($F_B$), as shown in Fig. 1a. Similar to Lorentz's reciprocity theorem, which governs transmission symmetry for electromagnetic wave phenomena, Maxwell-Betti's theorem stems from time-reversal symmetry and the principle of microscopic reversibility[22], and it is thus widely applicable. Breaking reciprocity in statics may significantly extend the functionalities of mechanical systems, in the same way that electromagnetic non-reciprocal devices[1,2,4], such as isolators[3] and circulators[5,6], have become essential components of modern electromagnetic systems. Maxwell-Betti's theorem is derived under a few basic assumptions, including the fact that the system under analysis is linear. We can therefore expect that Eq. 1 may be broken in suitably designed non-linear systems. In the following, we explore this path to induce strong static non-reciprocity for moderate applied force intensities using mechanical metamaterials[7-16]. We further harness their tunable, nonlinear and topological responses to induce giant mechanical isolation for applied static forces of moderate magnitude.

Consider first the "fishbone" mechanical metamaterial shown in Fig. 1b consisting of horizontal and transverse elastic beams, with the latter clamped to the laboratory frame and tilted by an angle $\theta$ with respect to the vertical axis. The angle $\theta$ quantifies the geometrical asymmetry of the structure, with $\theta = 0$ corresponding to a symmetric device. Fig. 1b shows the experimentally observed deformation of such a structure for an applied force $F_A = F_0 = -1N$ ($F_B = -F_0 = 1N$) at its left (right) side —see also Supplementary Video 1. We clearly see by naked eye that the output displacement is much larger for excitation from the left-hand side than from the right-hand side $u_{B \to A} \ll -u_{A \to B}$, evidence of a strongly non-reciprocal transmission of the displacement field. Such strong asymmetry results from the instability of the fishbone metamaterial—its abrupt switching into a different configuration for a given applied intensity—induced when excited from its left end, and associated to its large non-linear response. As shown by our measurements and numerical simulations (Fig. 1c), the instability onset corresponds to the point where the slope of the red curve suddenly becomes very large ($F_0 \approx -0.3N$). A comparable instability point also occurs when we *push* the material from the right end ($F_0 \approx 0.3N$). Such large non-reciprocal response induces up to 20 dB mechanical isolation (Fig. 1d), defined as the ratio of displacements $20 \log_{10} u_{A \to B}/-u_{B \to A}$.

Further insight into the non-reciprocal properties of the structure can be obtained by studying the non-reciprocity parameter $\Delta u \equiv u_{A \to B} + u_{B \to A}$, which is plotted in Fig. 1e for three asymmetry angles, $\theta = 0$, $\theta = \pi/16$ and $\theta = \pi/8$. For $\theta = 0$, symmetry requires $\Delta u = 0$, no matter how large the non-linearity is. In asymmetric structures, however, $\Delta u$ has a quadratic dependence on $F_0$ for small input forces, which is consistent with the fact that the leading linear terms of the Taylor expansion of $u_{A \to B}$ and $u_{B \to A}$ with respect to $F_0$ should satisfy Eq. (1). The figure shows that, although the structure with $\theta = \pi/8$ is more asymmetric than the one with $\theta = \pi/16$, the latter exhibits a much larger $\Delta u$. These findings suggest that strong static non-reciprocity stems from a delicate balance between asymmetry and nonlinearity, a fact that will become clearer in the following discussion.

The unusual relation between $\Delta u$ and $\theta$ in the structure of Fig. 1 can be understood through the simplified model in Fig. 2a, consisting of discrete nodes connected together through linear and torsional springs, modelling the beams and connections between them, respectively. In the Supplementary Information, we demonstrate that such a structure can be described through the equation

$$0 = \frac{d^2 u}{dx^2} - \lambda_1 u - \lambda_2 u^2 - \cdots, \qquad (2)$$

where $u(x)$ is the horizontal displacement of the central nodes and $\lambda_i$ are parameters that depend on $\theta$ and the spring properties. The solution of this equation is presented in Figs. 2ab for $\theta = \pi/16$ without (dashed) and with (solid) the non-linear terms in Eq. (2), for excitation from the left and right end, respectively, and input force $F_0 = -2N$. Despite the geometrical asymmetry, in the linear case the displacement fields $u(x)$ for excitations from opposite directions are identical, since the linearized version of Eq. (2) is inversion-symmetric, *i.e.,* invariant under the transformation $x \to -x$ and $u \to -u$. It follows that $\Delta u = 0$, as expected from Maxwell-Betti's theorem. In contrast, the nonlinear terms of even order in Eq. (2) break inversion-symmetry and induce asymmetric displacement profiles $u(x)$ for excitation from opposite sides, which subsequently results in $\Delta u \neq 0$. By solving Eq. (2), it is possible to prove (see Supplementary Information) that $\Delta u = \kappa(\theta) F_0^2$ for small $F_0$, where $\kappa$ is a $\theta$-dependent parameter associated with the structure, validating our prediction about the quadratic dependence of $\Delta u$ versus $F_0$ derived above through general principles. Despite the apparent simplicity of this model, such prediction (Fig. 2c) is in good qualitative agreement with experiments and simulations (Fig. 1e). Interestingly, the calculated non-reciprocal susceptibility $\kappa$ behaves non-monotically with the asymmetry angle $\theta$ (Fig. 2d), consistent with our observations from Fig. 1e: for a symmetrical structure $\theta \to 0$, $\kappa(\theta)$ vanishes, as expected, and for a large asymmetry $|\theta| \geq \pi/4$, the susceptibility also tends to zero, because non-linear effects become weaker. As a result of such interplay between non-linearity and asymmetry, $\kappa(\theta)$ exhibits two extrema for which the non-reciprocity is the strongest. Their position is directly controlled by the properties of the constitutive springs.

An important question to address is how to engineer large values of the susceptibility $\kappa$, in order to induce giant non-reciprocal responses for small input forces. A straightforward approach would be to use constitutive materials with smaller elastic moduli, resulting in overall smaller input forces, but in many practical cases we are restricted in terms of available material properties. An alternative strategy consists in realizing metamaterials with a large structural compliance. We do so by replacing the linear spring elements of the fishbone (Figs. 2ab) by freely hinging squares and bars (Fig. 2ef). The transverse bars play the same role as in the fishbone, and their tilt angle $\theta$ quantifies the structure asymmetry. As we discuss in the Supplementary Information, this structure is a mechanism, *i.e.*, it supports a single internal hinging motion, where all squares and bars pivot harmoniously. This mechanism costs elastic energy, which we model using torsional springs located at the central nodes. For a small input force $F_0 = -0.15N$, we observe a strongly asymmetric response, which *amplifies* (resp. *curbs*) the displacement when actuated from the left (resp. right) end (Fig. 2ef), in contrast with the fishbone structure, in which the displacement fields always decay from the excitation point. While the metamaterial still exhibits the same quadratic scaling with input intensity $\Delta u = \kappa(\theta)F_0^2$ (Fig. 2g), as expected from Eq. (1), the susceptibility $\kappa$ is several orders of magnitude larger than for the fishbone structure, and its dependence with the asymmetry angle $\theta$ is significantly different: $\kappa(\theta)$ diverges at $\theta \to 0$ and it monotonically decreases away from this singular point (Fig. 2h).

Considering that $\theta \to 0$ is the limit of a spatially symmetric structure, the divergence of the non-reciprocal parameter $\kappa$ is particularly interesting. As shown in the Supplementary Information, to linear order the displacement profile takes the form $u(n) = u(0)[g(\theta)]^n$, where $g(\theta)$ is a characteristic constant of the structure. In the limit of zero torsional stiffness, the structure is isostatic and supports this displacement distribution at zero energy, directly stemming from its nontrivial topological properties, akin to those found in topological mechanical metamaterials at zero frequency[23-26]. It can be indeed shown (see Supplementary Information) that the constant $g(\theta)$ is associated to the topological invariant $W$, defined in agreement with Ref. 23 and discussed in the Methods. The zero-energy displacement distribution is exponentially localized on the left (right) of the structure, associated with $W=0$ ($W=1$) for positive (negative) values of $\theta$, with a decay length proportional to $1/\log g(\theta)$. In the limit of a symmetric configuration ($\theta = 0$), this topologically-induced edge mode is delocalized, with its decay length going to infinity, while the topological invariant exhibits a discontinuity, consistent with the observations in Ref. 23. Around this topological transition the mode delocalization strongly connects the two metamaterial edges, leading to the divergence of the non-reciprocal susceptibility $\kappa$, and implying that the condition $\theta \simeq 0$ is ideally suited to enable a strong static non-reciprocal response for small applied forces. In contrast, the fishbone structure can be proven to be topologically trivial in the low-intensity regime—see Supplementary Information and Methods—leading to a larger threshold to observe non-reciprocal responses, enabled by the nonlinearities for larger forces. These results indicate that suitably tailored asymmetry *and* nonlinearity are the two fundamental requirements to achieve non-reciprocity in statics. Topological order is not necessary to realize this effect, but it provides an efficient framework to largely enhance it.

Inspired by the architecture in Figs. 2ef, we designed and built a 2D topological mechanical metamaterial (see Methods) showing non-reciprocity for small input forces, as well as displacement amplification (Fig. 3a; Supplementary Video 2). Such amplification occurs when the structure is actuated from its left end (red curve in Fig. 3b). In contrast, the response to actuation from its right end induces a decreasing displacement field (blue curve in Fig. 3b). Such difference builds up a strong non-reciprocal response, which can be probed both in simulations and in experiments (Fig. 3c), and that is qualitatively similar to the model discussed above. Since topological properties are inherently robust to continuous perturbations, the overall non-reciprocal response of the designed metamaterial is expected to be robust to continuous changes, defects and imperfections. The use of instabilities controlled by confinement[11] or by the texture of the probe[16] further provide a promising strategy to change these topological properties in a controlled fashion, achieving tunable topological and non-reciprocal responses.

In contrast to wave-based systems, which are inherently dynamic, a vast number of mechanical systems primarily operate in the static regime. In this Letter, we have shown how nonreciprocal metamaterials with large nonlinearities and suitably tailored asymmetries combined with topologically non-trivial features can support strong non-reciprocity and isolation for static applied forces. These systems open unique opportunities for devices with unprecedented static one-way functionalities, of relevance for shock and low-frequency vibration damping[15,27,28], mechanical energy harvesting[29], prosthetics and opto-mechanics[30].

# METHODS

**Sample geometries.** The first chain is 20 mm wide, 8mm high and 120 mm long, and is made of 10 repeatedly stacked unit cells (See Extended Data Figure 1a). The thickness of the horizontal beam (respectively transverse beams) is 2 mm (respectively 1 mm). We control the asymmetry by imposing the angle $\theta$ between the struts and the vertical axis, and investigate three structures where $\theta = 0$, $\pi/16$ and $\pi/8$.

The second metamaterial is 75mm wide, 8mm high and 120mm long, is made of 4 repeatedly stacked unit cells (See Extended Data Figure 1b). The unit cell is made of squares of diagonal length 10mm and diamond shaped quadrilaterals of diagonals lengths 16mm and 8 mm. The thickness of the connections between the vertices of these shape is 1mm.

**Sample Fabrication.** We create our structures by casting a two component silicon rubber with well-calibrated elastic properties (PVS, Zhermarck, Elite Double 32, Young's modulus 1.2 MPa) into 3D printed moulds. We let the silicon rubber cure for a few hours, after which we extract the sample by breaking the mould apart. We wait for one week after which the elastic properties of the polymer have settled. Finally, we glue the samples edges onto an aluminium frame, which allows us to confine the sample laterally.

**Mechanical testing and data acquisition.** In order to minimize alignment bias, we carefully align our structures within a 0.1mm accuracy in the frame. We position such frame in a uniaxial testing device (Instron 3366), which we equip with a 10 N load cell and which allows us to impose the input position better than 0.01 mm and to measure the input force $F_0$ better than 0.5 mN. The samples are probed mechanically by custom made mechanical tweezers. We subsequently apply point forces at points *(A)* and *(B)* (as in Fig. 1a) simply by changing the orientation of the frame with respect to the tensile tester. In parallel, we measure the output displacement by using a high-resolution camera (Basler 3840px x 2748px) synchronized with the tensile tester and use a custom made sub-pixel detection algorithm that allows us to track the output displacement within a 0.002 mm accuracy. In order to optimize the contrast of the image acquisition between the structures and the background, black opaque fabric has been glued onto the supporting aluminium frames.

**Numerical Protocol.** For our static finite elements simulations, we use the commercial software Abaqus/Standard and we use a neohookean energy density as a material model, using a shear modulus, $G$ = 0.40 MPa and bulk modulus, K=20.0 GPa) in plane strain conditions with hybrid quadratic triangular elements (abaqus type CPE6H). We perform a mesh refinement study in order to ensure that the thinnest parts of the samples where most of the stress and strain localized are meshed with at least four elements. As a result, the two metamaterials approximately have $2 \times 10^4$ triangular elements. We apply boundary conditions that correspond closely to the experiments.

**Topological Properties.** The fishbone metamaterial has a topologically trivial static linear response, which is symmetric despite of its structural asymmetry. This property is confirmed by the fact that its lowest-order modes have a finite non-zero energy, and that the phonon spectrum always has a gap, see Supplementary Information for more details. In contrast, our second (topological) metamaterial, similarly to earlier examples in the literature[23-25,31-34], is isostatic—its number of degrees of freedom equals its geometric constraints—and it supports a single edge mode, a localized soft mode of deformation at the edge. The response of such a structure is characterized by a topological invariant, the winding number $W$, which takes different values as $\theta$ changes sign, as *W=0* for $\theta > 0$ and *W=1* for $\theta < 0$, and which determines whether the edge mode is localised on the right (for *W=0*) or on the left (for *W=1*). The existence of a topological invariant in such an isostatic mechanical system is intimately related to the absence of inversion-symmetry in the linear response. In addition, such property indicates that the system is topologically protected, namely the nature of its response—here the side where the edge mode is localised—is robust to continuous perturbations of the structure, provided that it does not largely perturb the asymmetry $\theta$, switching the integer topological invariant $W$.

# REFERENCES

1	Potton, R. J. Reciprocity in optics. *Rep. Prog. Phys.* **67**, 717-754, doi:Pii S0034-4885(04)26328-X10.1088/0034-4885/67/5/R03 (2004).

2	Lira, H., Yu, Z., Fan, S. & Lipson, M. Electrically driven nonreciprocity induced by interband photonic transition on a silicon chip. *Phys. Rev. Lett.* **109**, 033901, doi:10.1103/PhysRevLett.109.033901 (2012).

3	Fan, L. *et al.* An all-silicon passive optical diode. *Science* **335**, 447-450, doi:10.1126/science.1214383 (2012).

4	Peng, B. *et al.* Parity–time-symmetric whispering-gallery microcavities. *Nat. Phys.* **10**, 394-398, doi:10.1038/Nphys2927 (2014).

5	Estep, N. A., Sounas, D. L., Soric, J. & Alù, A. Magnetic-free non-reciprocity and isolation based on parametrically modulated coupled-resonator loops. *Nat. Phys.* **10**, 923-927 (2014).

6	Fleury, R., Sounas, D. L., Sieck, C. F., Haberman, M. R. & Alu, A. Sound isolation and giant linear nonreciprocity in a compact acoustic circulator. *Science* **343**, 516-519, doi:10.1126/science.1246957 (2014).

7	Lakes, R. Foam structures with a negative Poisson's ratio. *Science* **235**, 1038-1040 (1987).

8      Mullin, T., Deschanel, S., Bertoldi, K. & Boyce, M. C. Pattern Transformation Triggered by Deformation. *Phys. Rev. Lett.* **99**, 084301 (2007).

9      Schaedler, T. A. *et al.* Ultralight metallic microlattices. *Science* **334**, 962-965, doi:10.1126/science.1211649 (2011).

10     Silverberg, J. L. *et al.* Using origami design principles to fold reprogrammable mechanical metamaterials. *Science* **345**, 647-650 (2014).

11     Florijn, B., Coulais, C. & van Hecke, M. Programmable Mechanical Metamaterials. *Phys. Rev. Lett.* **113**, doi:10.1103/PhysRevLett.113.175503 (2014).

12     Nash, L. M. *et al.* Topological mechanics of gyroscopic metamaterials. *Proc. Natl. Ac. Sc. U. S. A.* **112**, 14495-14500, doi:10.1073/pnas.1507413112 (2015).

13     Shan, S. *et al.* Multistable Architected Materials for Trapping Elastic Strain Energy. *Adv. Mater.* **27**, 4296-4301, doi:10.1002/adma.201501708 (2015).

14     Susstrunk, R. & Huber, S. D. Observation of phononic helical edge states in a mechanical topological insulator. *Science* **349**, 47-50, doi:10.1126/science.aab0239 (2015).

15     Frenzel, T., Findeisen, C., Kadic, M., Gumbsch, P. & Wegener, M. Tailored Buckling Microlattices as Reusable Light-Weight Shock Absorbers. *Adv. Mater.*, doi:10.1002/adma.201600610 (2016).

16     Coulais, C., Teomy, E., de Reus, K., Shokef, Y. & van Hecke, M. Combinatorial Design of Textured Mechanical Metamaterials. *Nature* **535**, 529-532, doi:10.1038/nature18960 (2016).

17     Maxwell, J. C. L. On the calculation of the equilibrium and stiffness of frames. *Philosophical Magazine Series 4* **27**, 294-299, doi:10.1080/14786446408643668 (1864).

18   Betti, E. Teoria della elasticita'. *Il Nuovo Cimento (1869-1876)* **7**, 69-97, doi:10.1007/BF02824597 (1872).

19   Charlton, T. M. A Historical Note on the Reciprocal Theorem and Theory of Statically Indeterminate Frameworks. *Nature* **187**, 231-232, doi:10.1038/187231a0 (1960).

20   Love, A. E. H. *A Treatise on the Mathematical Theory of Elasticity*. (Cambridge University Press, 2013).

21   Timoshenko, S. & Young, D. H. *Theory of structures*. (McGraw-Hill, 1965).

22   Casimir, H. B. G. On Onsager's principle of microscopic reversibility. *Rev. Mod. Phys.* **17**, 343 (1945).

23   Kane, C. L. & Lubensky, T. C. Topological boundary modes in isostatic lattices. *Nat. Phys.* **10**, 39-45 (2014).

24   Chen, B. G., Upadhyaya, N. & Vitelli, V. Nonlinear conduction via solitons in a topological mechanical insulator. *Proc. Natl. Ac. Sc. U. S. A.* **111**, 13004-13009, doi:10.1073/pnas.1405969111 (2014).

25   Chen, B. G. *et al.* Topological Mechanics of Origami and Kirigami. *Phys. Rev. Lett.* **116**, 135501, doi:10.1103/PhysRevLett.116.135501 (2016).

26   Huber, S. D. Topological mechanics. *Nat. Phys.* **12**, 621-623, doi:10.1038/nphys3801 (2016).

27   Kadic, M., Bückmann, T., Schittny, R. & Wegener, M. Metamaterials beyond electromagnetism. *Rep. Prog. Phys.* **76**, 126501 (2013).

28   Brule, S., Javelaud, E. H., Enoch, S. & Guenneau, S. Experiments on seismic metamaterials: molding surface waves. *Phys. Rev. Lett.* **112**, 133901, doi:10.1103/PhysRevLett.112.133901 (2014).

29  Hussein, M. I., Leamy, M. J. & Ruzzene, M. Dynamics of Phononic Materials and Structures: Historical Origins, Recent Progress, and Future Outlook. *Appl. Mech. Rev.* **66**, 040802-040802, doi:10.1115/1.4026911 (2014).

30  Aspelmeyer, M., Kippenberg, T. J. & Marquardt, F. Cavity optomechanics. *Rev. Mod. Phys.* **86**, 1391-1452 (2014).

31  Paulose, J., ge Chen, B. G. & Vitelli, V. Topological modes bound to dislocations in mechanical metamaterials. *Nat Phys* **11**, 153-156, doi:10.1038/nphys3185 (2015).

32  Lubensky, T. C., Kane, C. L., Mao, X., Souslov, A. & Sun, K. Phonons and elasticity in critically coordinated lattices. *Rep. Prog. Phys.* **78**, 073901, doi:10.1088/0034-4885/78/7/073901 (2015).

33  Paulose, J., Meeussen, A. S. & Vitelli, V. Selective buckling via states of self-stress in topological metamaterials. *Proc Natl Acad Sci U S A* **112**, 7639-7644, doi:10.1073/pnas.1502939112 (2015).

34  Meeussen, A. S., Paulose, J. & Vitelli, V. Geared topological metamaterials with tunable mechanical stability. *Phys. Rev. X* **6**, 041029 (2016).

## ACKNOWLEDGEMENTS

We thank D. Ursem for his skilful technical assistance. We are grateful to M. van Hecke, V. Vitelli, A. Souslov, Y. Hadad, A. Meeussen and S. Waitukaitis for insightful discussions. C.C. acknowledges funding from the Netherlands Organization for Scientific research (NWO) VENI grant No. NWO-680-47-445. D.S. and A.A. were supported by the Air Force Office of Scientific Research, the Office of Naval Research, and the Simons Foundation.

## AUTHORS CONTRIBUTIONS

C.C., D.S. and A.A. developed the concepts. C.C. performed the experiments and the numerical simulations. C.C. and D.S. carried out the theoretical analysis. C.C., D.S. and A.A. wrote the paper.

**FIGURES**

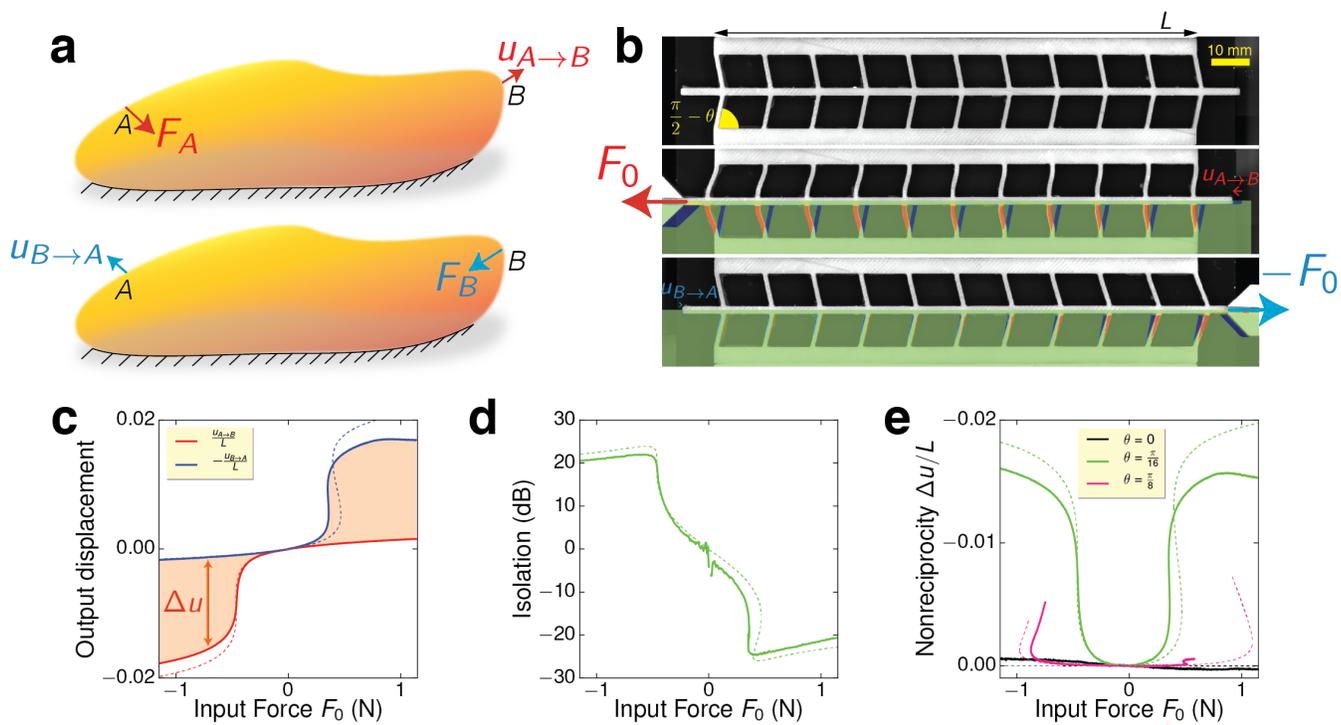

**Figure 1. Static non-reciprocity**. a. A structure is subject to a point force $F_A$ at point $A$ ($F_B$ at point $B$), inducing a displacement $u_{A \to B}$ at point $B$ ($u_{B \to A}$ at point $A$). b. 3D printed asymmetric nonlinear geometry actuated with $F_A = F_0$ and $F_B = -F_0$ at its two ends with $F_0 = -1\,N$, showing a strong non-reciprocal response $u_{B \to A} \ll -u_{A \to B}$ (see Methods for fabrication and experimental details). For clarity, the image difference between the deformed and initial geometries have been overlaid on the bottom half of the pictures. c. Rescaled output displacement $u_{A \to B}/L$ (red) and $-u_{B \to A}/L$ (blue) vs. input force $F_0$ (numerical simulations, dashed; experiments, solid). The shaded area represents $\Delta u = u_{B \to A} + u_{A \to B}$. d. Isolation vs. input force $F_0$. e. Non-reciprocity parameter $\Delta u$ vs. input force $F_0$ for different values $\theta = 0$ (black), $\theta = \pi/16$ (green) and $\theta = \pi/8$ (pink) (simulations, dashed; experiment, solid).

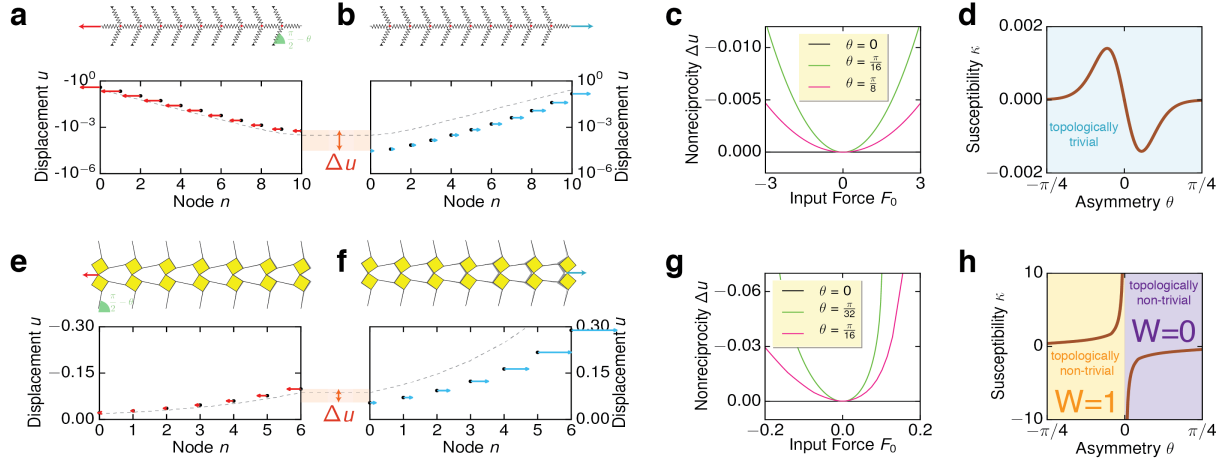

**Figure 2. Discrete Models for Non-Reciprocal Metamaterials.** ab. Blueprint of the fishbone structure considered in Fig. 1. Corresponding displacement fields $u$ vs. $n$ for excitations from the left end (a) and the right end (b) for an input force $F_0 = -2\,N$. The dashed lines correspond to the linearized problem. The shaded area represents $\Delta u = u_{B \to A} + u_{A \to B}$. c. Non-reciprocity $\Delta u$ vs. input force $F_0$ for different values of $\theta$. d. Non-reciprocal susceptibility $\kappa$ vs. the asymmetry $\theta$. The metamaterial is topologically trivial. ef. Blueprint of the topological mechanical metamaterial, we display both the undeformed (gray) and deformed structures (yellow) and the corresponding displacement fields $u$ vs. $n$ for excitations from the left end (e) and the right end (f) for an input force $F_0 = -0.15\,N$. The dashed lines correspond to the linearized problem. g. Non-reciprocity parameter $\Delta u$ vs. input force $F_0$ for different values of the asymmetry $\theta$. h. Non-reciprocal susceptibility $\kappa$ and topological invariant (called the winding number) W vs. the asymmetry $\theta$. The metamaterial is topologically non-trivial and its winding number equals 0 (1) for $\theta > 0$ ($\theta < 0$).

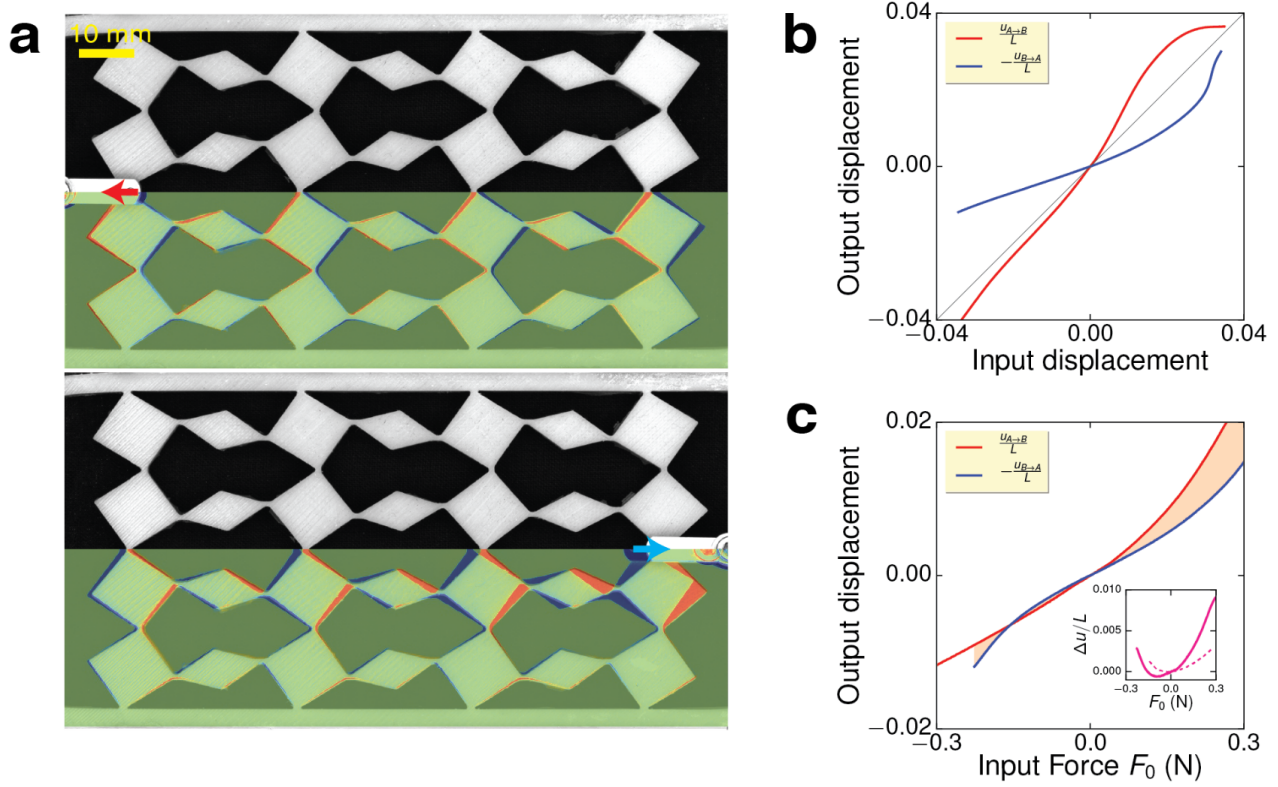

**Figure 3. 2D Topological Mechanical Metamaterial.** a. Metamaterial with a microstructure inspired by Fig. 2ef, upon excitation from the left end (top) and right end (bottom) for an input force $F_0 = -0.2\ N$. For clarity, the image difference between the deformed and initial geometries have been overlaid on the bottom half of the pictures. bc. Corresponding output displacement $u_{A \to B}/L$ (red) and $-u_{B \to A}/L$ (blue) vs. input displacement $u_0$ (b) and input force $F_0$ (c). The shaded area corresponds to $\Delta u = u_{B \to A} + u_{A \to B}$. c-Inset. $\Delta u$ vs. input force $F_0$ for simulations (dashed) and experiments (solid).

**EXTENDED DATA**

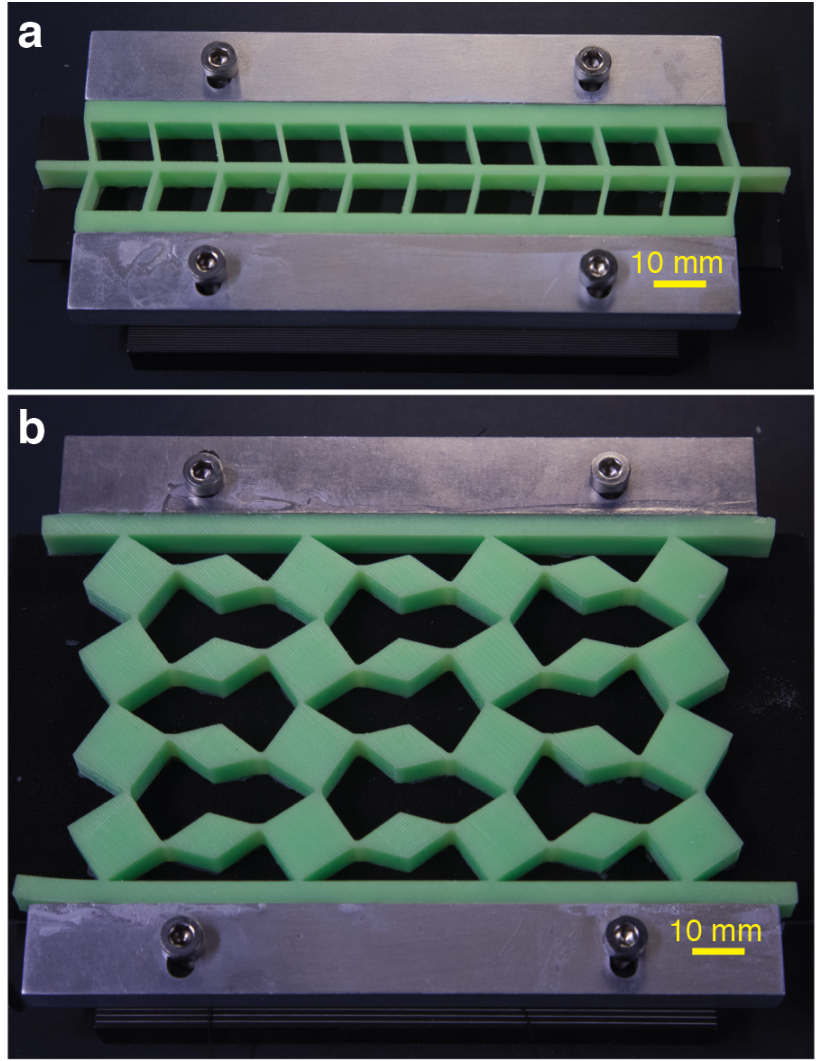

**Extended Data Figure 1. Pictures of the mechanical Metamaterials in their confining frames.** a. Fishbone (a) and topological (b) mechanical metamaterials, both with an asymmetry angle $\theta = \pi/16$.

# Supplementary Information for the paper *Static Non-reciprocity in Mechanical Metamaterials*

Corentin Coulais,[1,2] Dimitrios Sounas,[3] and Andrea Alù[1,3]

[1]*FOM Institute AMOLF, Science Park 104, 1098 XG Amsterdam, The Netherlands*
[2]*Huygens-Kamerlingh Onnes Lab, Universiteit Leiden, PObox 9504, 2300 RA Leiden, The Netherlands*
[3]*Department of Electrical and Computer Engineering,*
*The University of Texas at Austin, Austin, Texas 78701, USA*

# I. DISCRETE MODEL FOR THE FISHBONE MECHANICAL METAMATERIAL

## A. Governing equation

The Lagrangian of the mechanical systems depicted in Fig. S1 is

$$\mathcal{L} = \sum_{n=1}^{N}\left[k_1\varepsilon_n^2 + \frac{C}{2}\delta\theta_n^2\right] + \sum_{n=0}^{N}\left[\frac{k_2}{2}(u_{n+1}-u_n)^2\right] \\ + \sum_{n=0}^{N+1}[f_n b\left((1+\varepsilon_n)\sin(\theta+\delta\theta_n) - \sin\theta - u_n\right) + g_n b\left((1+\varepsilon_n)\cos(\theta+\delta\theta_n) - \cos\theta\right)] \quad (1)$$

where $f_n$ and $g_n$ are Lagrange multipliers. They originate from the fact that (i) the chain is symmetric with respect to the horizontal axis ; (ii) we confine the chain in the vertical direction. When the system is at mechanical equilibrium, the Lagrangian remains unchanged upon small variations of the internal degrees of freedom. Therefore, for every $n \in [0, N]$, $\partial\mathcal{L}/\partial\delta\theta_n = 0$, $\partial\mathcal{L}/\partial\varepsilon_n = 0$ and $\partial\mathcal{L}/\partial u_n = 0$. After eliminating $f_n$ and $g_n$ from the resulting equations, we obtain a system of nonlinear coupled equations

$$0 = k_2(2u_n - u_{n+1} - u_{n-1}) + 2bk_1\varepsilon_n \sin(\theta+\delta\theta_n) + Cb\frac{\delta\theta_n}{1+\varepsilon_n}\cos(\theta+\delta\theta_n) \quad \text{for} \quad n \in [1, N] \quad (2)$$

$$0 = k_2(u_0 - u_1) - f_0 \quad (3)$$

$$0 = k_2(u_{N+1} - u_N) - f_{N+1}, \quad (4)$$

where the opposite of Lagrange multipliers $-f_0$ and $-f_{N+1}$ correspond to the reaction forces exerted at the left and right edges of the chain, respectively. In the following, we will express the mechanics of the chains as a function of the horizontal displacement $u_n$. To do this, first recall that the geometrical constraints can be expressed as follows

$$u_n = b\left((1+\varepsilon_n)\sin(\theta+\delta\theta_n) - \sin\theta\right) \quad (5)$$

$$0 = (1+\varepsilon_n)\cos(\theta+\delta\theta_n) - \cos\theta, \quad (6)$$

for $n \in [0, N+1]$. Assuming that $\delta\theta_n \ll \theta$, that $\varepsilon_n \ll 1$ and that $u_n \ll 1$, we can expand the above equation to second order in $\delta\theta_n$, $\varepsilon_n$ and $u_n$ and we can substitute the previous system of coupled nonlinear equations into a system of

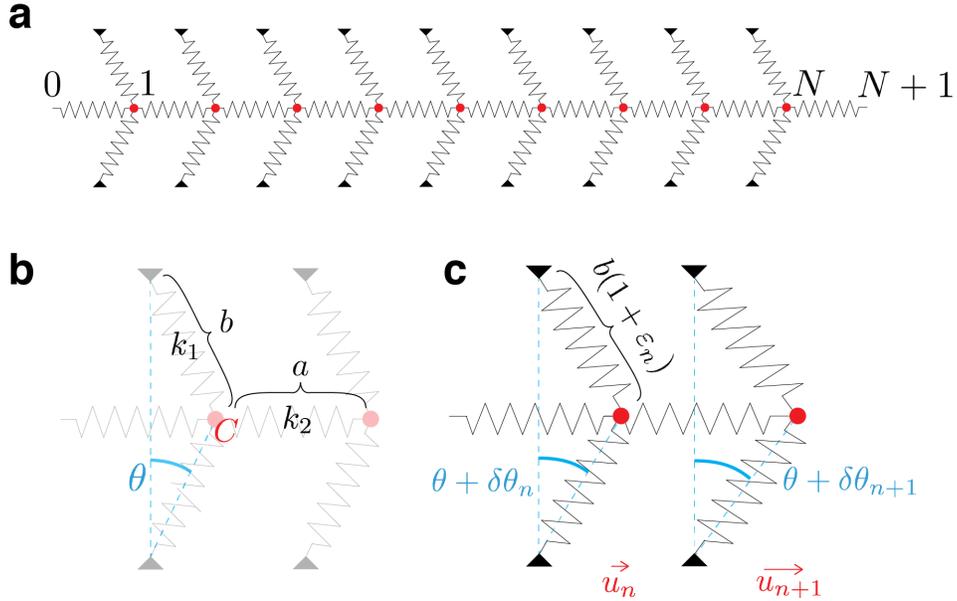

FIGURE S1. Sketch of the fishbone mechanical metamaterial. a. Geometry of the chain constituted of $N+2$ nodes attached to linear horizontal springs (stiffness $k_2$), to transverse springs of stiffness $k_1$, to torsional springs $C$ and with an initial tilt angle $\theta$. bc. Geometry of two unit cells in undeformed (b) and deformed (c) configurations. The data shown in figures 2a-d of the main text has been produced using $a = 1$, $b = 1$, $k_1 = 2$, $k_2 = 1$ and $C = 1$.

quadratic equations in $u_n$, for $n \in [1, N]$

$$k_2(2u_n - u_{n+1} - u_{n-1}) + \left(k_1(1 - \cos(2\theta)) + \frac{C}{2}(1 + \cos(2\theta))\right) u_n + 3(k_1 - C)b^{-1} \sin(\theta) \cos^2(\theta) u_n^2 = 0, \quad (7)$$

In addition, the forces exerted at boundary nodes $n = 0$ and $n = N + 1$ are expressed as follows $F_0 = -k2(u_0 - u_1)$ and $F_{N+1} = -k2(u_{N+1} - u_N)$. The boundary conditions $(A)$ (respectively $(B)$) correspond to applying a point force of magnitude $F_0$ on the node $0$ (resp. $N + 1$) and thus read $F_{0(A)} = F_0$ and $F_{N+1(A)} = 0$ (respectively $F_{0(B)} = 0$ and $F_{N+1(B)} = -F_0$).

### B. Non-reciprocal properties

In this section, we will solve Eq. 7 and determine the non-reciprocal properties of the chain. To do this, we take the continuum limit assuming $a \to dx$, $u_n \to u(x)$ and taking a Taylor series around $x = n$, $u_{n-1} \to u(x) - a(du/dx) + a^2/2(d^2u/dx^2)$ and $u_{n+1} \to u(x) + a(du/dx) + a^2/2(d^2u/dx^2)$. Next, we perform a change of units $x \to x/a$ and obtain

$$\frac{d^2u}{dx^2} - \lambda_1 u - \lambda_2 u^2 = 0, \quad (8)$$

with

$$\lambda_1 = \frac{1}{2k_2 a^2}\left(k_1(1 - \cos(2\theta)) + \frac{C}{2}(1 + \cos(2\theta))\right) \text{ and } \lambda_2 = \frac{1}{2k_2 a^2} 3(k_1 - C)b^{-1}\sin(\theta)\cos^2(\theta), \quad (9)$$

and

$$F_0 = k_2 a \left(\frac{du}{dx}\right)_{x=0} \text{ and } F_{N+1} = -k_2 a \left(\frac{du}{dx}\right)_{x=N+1}. \quad (10)$$

In particular, assuming that the input force is $F_0$, we express the solution to Eq. (7) as

$$u(x) = F_0 \, s^{(1)}(x) + F_0^2 \, s^{(2)}(x) + \ldots \quad (11)$$

where $s^{(1)}(x)$ (respectively $s^{(2)}(x)$) is the local compliance at first order (respectively second order). Within this perturbative expansion, boundary condition $(A)$ translates into

$$\frac{ds^{(1)}}{dx}(x = 0) = \frac{1}{k_2 a}, \; \frac{ds^{(1)}}{dx}(x = N + 1) = 0, \; \frac{ds^{(2)}}{dx}(x = 0) = 0 \text{ and } \frac{ds^{(2)}}{dx}(x = N + 1) = 0. \quad (12)$$

To first order in $F_0$, Eq. (8) is

$$0 = \frac{d^2 s^{(1)}}{dx^2} - \lambda_1 s^{(1)} \quad (13)$$

and to second order, we have

$$0 = \frac{d^2 s^{(2)}}{dx^2} - \lambda_1 s^{(2)} - \lambda_2 s^{(1)^2} \quad (14)$$

Solving these two equations for boundary condition $(A)$,

$$s^{(0)}(x) = -\frac{1}{k_2 a \sqrt{\lambda_1}} \frac{\cosh\left(\sqrt{\lambda_1}(N' - x)\right)}{\sinh\left(\sqrt{\lambda_1} N'\right)} \quad (15)$$

and

$$s^{(1)}(x) = \frac{\lambda_2 e^{\sqrt{\lambda_1} x}}{3k_2^2 a^2 \lambda_1^2 (e^{2\sqrt{\lambda_1} N'} - 1)^2} \Big(-2e^{2\sqrt{\lambda_1} N'} - 2e^{4\sqrt{\lambda_1} N'} \\ + e^{\sqrt{\lambda_1}(4N' - x)} - 2e^{2\sqrt{\lambda_1} x} + e^{3\sqrt{\lambda_1} x} - 2e^{2\sqrt{\lambda_1}(N' + x)} - 6e^{\sqrt{\lambda_1}(2N' + x)}\Big), \quad (16)$$

where $N' = N + 1$. To solve the equation (8) under boundary condition $(B)$, we notice that the change of variable $x \to x' = N' - x$ only changes the sign of the linear terms in eq. (8). Therefore, $u(x')_{(B)} = -F_0 s^{(1)}(x') + F_0^2 s^{(2)}(x')$. Finally, from the output displacements $u_{A \to B} \equiv u(N')_{(A)}$ and $u_{B \to A} \equiv u(0)_{(B)}$ we can calculate the non-reciprocal parameter discussed in the main text

$$\Delta u \equiv u_{A \to B} + u_{B \to A} = -\frac{2\lambda_2 F_0^2}{3k_2^2 a^2 \lambda_1^2} \frac{1 + 2\cosh(\sqrt{\lambda_1} N')}{\sinh^2 \sqrt{\lambda_1} N'}. \quad (17)$$

## C. Trivial static topological properties

We saw in the previous section that the static linear response of the fishbone metamaterial is symmetric and consistent with Maxwell-Betti's theorem. Therefore, we expect that the static topological properties—which are directly associated to the asymmetry of the linear response—to be trivial. In this section, we demonstrate this point explicitly by analysing the band structure and performing a mode analysis. To this end, we consider the linear version of Eq. (7)

$$0 = k_2(2u_n - u_{n+1} - u_{n-1}) + \lambda u_n, \tag{18}$$

where

$$\lambda = k_1(1 - \cos(2\theta)) + \frac{C}{2}(1 + \cos(2\theta)), \tag{19}$$

and for $1 < n < N$. We assume in the following that the end points of the metamaterial can freely move. Therefore, the Eqs. (3)-(4) describing the end points of the $k_2(u_0 - u_1) = 0$ and $k_2(u_{N+1} - u_N) = 0$. These equations can be recast in the matrix form

$$\underbrace{\begin{bmatrix} 0 \\ 0 \\ \vdots \\ \vdots \\ 0 \\ \vdots \\ 0 \end{bmatrix}}_{1 \times (N+2)} = \underbrace{\begin{bmatrix} u_0 \\ u_1 \\ \vdots \\ \vdots \\ u_n \\ \vdots \\ u_{N+1} \end{bmatrix}}_{1 \times (N+2)} \underbrace{\begin{bmatrix} k_2 & -k_2 & 0 & \cdots & \cdots & \cdots & \cdots & 0 \\ -k_2 & 2k_2+\lambda & -k_2 & 0 & \cdots & \cdots & \cdots & 0 \\ 0 & -k_2 & 2k_2+\lambda & -k_2 & 0 & \cdots & \cdots & 0 \\ 0 & 0 & \ddots & \ddots & \ddots & \ddots & \vdots & \vdots \\ \vdots & \vdots & & & -k_2 & 2k_2+\lambda & -k_2 & 0 \\ \vdots & \vdots & \cdots & \cdots & 0 & -k_2 & 2k_2+\lambda & -k_2 \\ 0 & 0 & \cdots & \cdots & \cdots & 0 & -k_2 & k_2 \end{bmatrix}}_{(N+2) \times (N+2)}. \tag{20}$$

The $(N+2) \times (N+2)$ matrix in the previous equation is called the stiffness matrix $\mathbf{K}$. For a structure where all the nodes have the same mass $m$, the dynamical matrix $\mathbf{D}$ is proportional to the stiffness matrix $\mathbf{D} = \mathbf{K}/m^1$. The vibrational modes of the metamaterial can be directly analysed from the spectral analysis of the dynamical matrix, where the eigenvalues are the square of the eigen-frequencies. The most straightforward method to investigate such vibrational properties is to assume at first the system infinite and perform a Fourier analysis. In an infinite system, the displacement field $u_n$ can be expressed as plane waves

$$u_n = \hat{u}(q)e^{iqn}, \tag{21}$$

As a result of the translation property of the Fourier transform, we obtain

$$0 = \hat{u}(q)D(q), \tag{22}$$

where

$$D(q) = 2\frac{k_2}{m} + \frac{\lambda}{m} - \frac{k_2}{m}e^{iq} - \frac{k_2}{m}e^{-iq} = \omega_0^2\left(2 + \frac{\lambda}{m\omega_0^2} - 2\cos q\right) \tag{23}$$

is a scalar that can be interpreted as the Fourier transform of the dynamical matrix. $\omega_0 = \sqrt{k_2/m}$ is the characteristic frequency of the structure. We plot in Fig. S2a the corresponding band diagram, which, noticeably, exhibits a band-gap $\Delta\omega$. The smoking gun of topologically non-trivial response in static structures is the presence of a vanishing band-gap in the phonon spectrum centered at zero-wave vector, which is often associated to a change of the topological properties[2]. Given that the parameter $\lambda$ remains always strictly positive for all values of the fishbone constitutive parameters, the dynamical matrix $D(q)$ never becomes zero. This demonstrates that the band-gap at zero wave vector $\Delta\omega$ never vanishes and shows that there is no link between the structural asymmetry and the topological invariant. This suggests that the system remains topologically trivial.

To prove this explicitly, we need to demonstrate the absence of zero-frequency edge modes. To do this, we perform such analysis numerically for a finite system size and we summarize our findings in Fig. S2b-d. This analysis reveals that the finite system has two edge modes of low, yet finite frequency (solid red and dotted blue lines in Fig. S2b and red bars in Fig. S2c), which, in contrast with the rest of the spectrum (gray lines in Fig. S2b and gray bars in Fig. S2c), are located in the band gap (depicted by the blue shaded area in Fig. S2c). Such edges modes are 1D equivalent of Rayleigh waves, which occur at the free-surface of elastic solids. We calculate the frequency of these edge modes for different structure asymmetries $\theta$ and notice that it is never zero and are therefore always "gapped" (see Fig. S2d). The absence of gapless edge modes is a direct proof that such structure is topologically trivial[3].

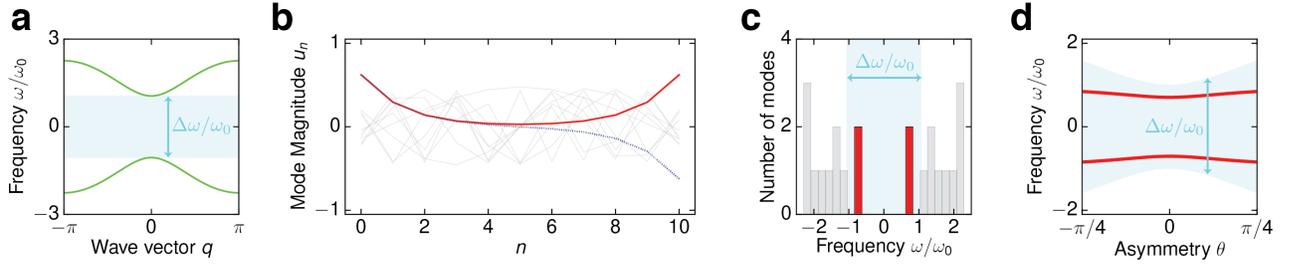

FIGURE S2. Mode analysis of the fishbone mechanical metamaterial. (a) Rescaled frequency $\omega/\omega_0$ vs. wave vector $q$ from the Fourier analysis carried out by assuming that the structure infinite. The blue shaded area denotes the band-gap. (b) Spatial profile of the eigenmodes of the finite structure ($N = 9$). The two eigenmodes with the lowest frequencies are displayed in solid red and dotted blue lines respectively and correspond to edge-modes. The higher frequency eigenmodes are displayed in solid light gray lines. (c) Histogram of the rescaled eigen-frequencies. The two edge-modes located in the band-gap (light blue shaded area) are shown in red and the remaining "bulk" modes in light gray. $\omega_0 = \sqrt{k_2/m}$ is the characteristic frequency of the structure. (d) Band-gap (light blue shaded area) and edge modes rescaled frequencies $\omega_E/\omega_0$ (thick red line) vs. asymmetry angle $\theta$ of the structure. All panels have been produced using $N = 9$, $a = 1$, $b = 1$, $k_1 = 2$, $C = 1$ and panels (a-c) have been produced using $\theta = \pi/16$.

## II. DISCRETE MODEL FOR THE MECHANICAL METAMATERIAL 2

### A. Kinematics

On the sketch depicted in figure S3, each unit cell has 4 nodes whose initial coordinates are

$$\mathbf{V_n} = 2\sqrt{2}na\cos\theta\mathbf{e_x} + a\sqrt{2}\left(\sin\theta\mathbf{e_x} + \cos\theta\mathbf{e_y}\right) \tag{24}$$

$$\mathbf{U_n} = \mathbf{V_n} + a\sqrt{2}\left(-\sin\theta\mathbf{e_x} + \cos\theta\mathbf{e_y}\right) \tag{25}$$

$$\mathbf{W_n} = \mathbf{V_n} + a\left(-\sin\left(\frac{\pi}{4}+\theta\right)\mathbf{e_x} + \cos\left(\frac{\pi}{4}+\theta\right)\mathbf{e_y}\right) \tag{26}$$

$$\mathbf{Z_n} = \mathbf{V_n} + a\left(-\sin\left(-\frac{\pi}{4}+\theta\right)\mathbf{e_x} + \cos\left(-\frac{\pi}{4}+\theta\right)\mathbf{e_y}\right), \tag{27}$$

where $b \equiv 2\sqrt{2}a\cos\theta$ is the lattice constant. We assume that the square formed by these four point is rigid.
Therefore, the deformed coordinates are

$$\mathbf{v_n} = 2\sqrt{2}na\cos\theta\mathbf{e_x} + a\sqrt{2}\left(\sin\left(\theta+\delta\phi_n\right)\mathbf{e_x} + \cos\left(\theta+\delta\phi_n\right)\mathbf{e_y}\right) \tag{28}$$

$$\mathbf{u_n} = \mathbf{v_n} + a\sqrt{2}\left(-\sin\left(\theta+\delta\theta_n\right)\mathbf{e_x} + \cos\left(\theta+\delta\theta_n\right)\mathbf{e_y}\right) \tag{29}$$

$$\mathbf{w_n} = \mathbf{v_n} + a\left(-\sin\left(\frac{\pi}{4}+\theta+\delta\theta_n\right)\mathbf{e_x} + \cos\left(\frac{\pi}{4}+\theta+\delta\theta_n\right)\mathbf{e_y}\right) \tag{30}$$

$$\mathbf{z_n} = \mathbf{v_n} + a\left(-\sin\left(-\frac{\pi}{4}+\theta+\delta\theta_n\right)\mathbf{e_x} + \cos\left(-\frac{\pi}{4}+\theta+\delta\theta_n\right)\mathbf{e_y}\right). \tag{31}$$

In addition, the link connecting the unit cell $n$ to $n+1$ is described by the vector

$$\ell_{\mathbf{n,n+1}} = a\sqrt{2}\left(\cos\left(-\theta+\delta\psi_{n,n+1}\right)\mathbf{e_x} + \sin\left(-\theta+\delta\psi_{n,n+1}\right)\mathbf{e_y}\right), \tag{32}$$

where $\delta\psi_{n,n+1}$ is the rotation angle of the bar connecting the sites $n$ and $n+1$. As a result, the deformation of each unit cell $n$ is described by 2 variables $\delta\phi_n$, $\delta\theta_n$, and the connection between unit cells $n$ and $n+1$ by the variable $\delta\psi_{n,n+1}$. Hence, a chain of length $N$ is fully described by $3N+2$ degrees of freedom. As subsequent unit cells are connected to one another, these variables are not independent, and are constrained by the following geometrical relation

$$\mathbf{w_{n+1}} = \mathbf{z_n} + \ell_{\mathbf{n,n+1}}, \tag{33}$$

that on a chain of length $N+1$ induces $2N$ geometrical constraints. Besides, as the chain is clamped and symmetric in the vertical axis, the vertical displacement of the vertex $U$ is zero, which enforces $N$ additional constraints. As there are $3N+2$ degrees of freedom and $3N+1$ geometrical constraints, the chain has one single degree of freedom. Therefore, the mechanics of the chain is a pure geometrical problem, which we solve by writing the aforementioned constraints using Eqs.(28-31) and obtaining

$$0 = a\sqrt{2}\left(\sin\left(\theta+\delta\phi_{n+1}\right)-\sin\left(\theta+\delta\phi_n\right)-\cos\left(-\theta+\delta\psi_{n,n+1}\right)\right)-a\left(\sin\left(\frac{\pi}{4}+\theta+\delta\theta_{n+1}\right)-\sin\left(-\frac{\pi}{4}+\theta+\delta\theta_n\right)\right)+b \quad (34)$$

$$0 = a\sqrt{2}\left(\cos\left(\theta+\delta\phi_{n+1}\right)-\cos\left(\theta+\delta\phi_n\right)-\sin\left(-\theta+\delta\psi_{n,n+1}\right)\right)+a\left(\cos\left(\frac{\pi}{4}+\theta+\delta\theta_{n+1}\right)-\cos\left(-\frac{\pi}{4}+\theta+\delta\theta_n\right)\right) \quad (35)$$

$$0 = \sqrt{2}\cos\left(\theta+\delta\theta_n\right) + \sqrt{2}\cos\left(\theta+\delta\phi_n\right) - 2\sqrt{2}\cos\theta. \quad (36)$$

As such mechanical system only has one degree of freedom, the kinematics is simply given by the solution to this systems of coupled nonlinear equations. We address the kinematics of this system in two ways : first, we consider small input deformations and express the contraints to first order in angular deflection ; second, we solve such system fully numerically.

### B. Governing equation

#### 1. Linearized governing equation : topological insulator

We perturb the chain mechanically with small input forces, therefore we may consider leading order nonlinearities and expand Equations (34-36) to first order. After substituting the angles $\delta\phi_n$ and $\delta\psi_{n,n+1}$, we obtain

$$0 = -\delta\theta_n(\sin(2\theta) + \cos(2\theta) + 2) + \delta\theta_{n+1}(-\sin(2\theta) + \cos(2\theta) + 2). \quad (37)$$

As the horizontal $u_n$ displacement is linearly related to the rotation angle $\delta\theta_n$,

$$u_n = (\mathbf{u_n} - \mathbf{U_n})\cdot\mathbf{e_x} = a\sqrt{2}(\sin(\theta+\delta\phi_n) - \sin(\theta+\delta\theta_n)) \approx -a2\sqrt{2}\cos(\theta)\delta\theta_n, \quad (38)$$

and Eq. (37 becomes

$$0 = -u_n(\sin(2\theta) + \cos(2\theta) + 2) + u_{n+1}(-\sin(2\theta) + \cos(2\theta) + 2). \quad (39)$$

This linear equation admits simple exponential solutions of the form

$$u_n = u_0 g(\theta)^n, \quad (40)$$

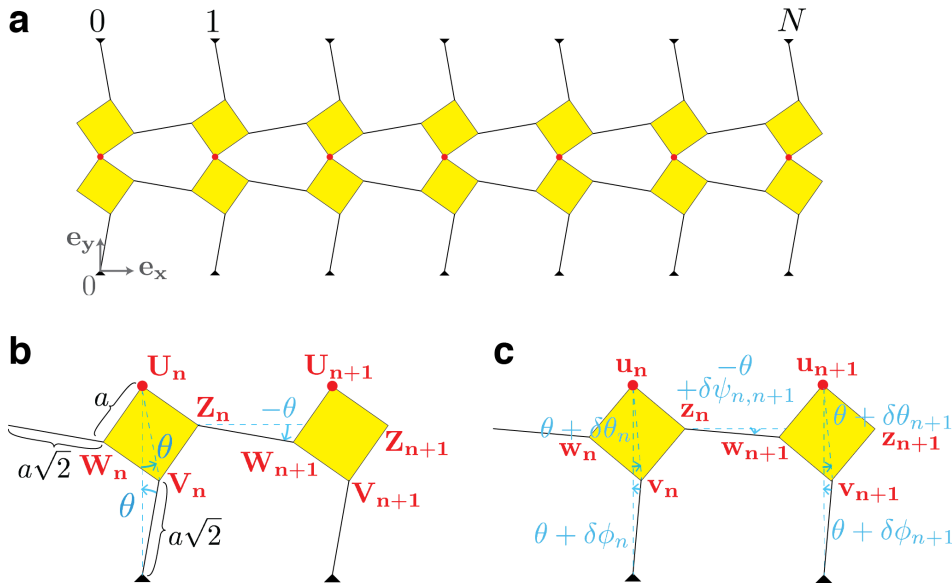

FIGURE S3. Sketch of the mechanical metamaterial 2. a. Geometry of the chain constituted of $N+1$ unit cells, characterized by their initial tilt angle $\theta$ and connected in the middle by a torsional spring $C$ (red dots). bc. Geometry of two unit cells in undeformed (b) and deformed (c) configurations. The data shown in figures 2e-h of the main text has been produced using $C = 1$.

where

$$g(\theta) = \left(\frac{2 + \cos(2\theta) + \sin(2\theta)}{2 + \cos(2\theta) - \sin(2\theta)}\right). \tag{41}$$

We notice that $g(0) = 1$, and that $g(\theta) > 1$ for $\theta > 0$ and $g(\theta) < 1$ for $\theta < 0$. Therefore, depending on the sign of $\theta$, the displacement field exhibits exponentially localized large values on the right side (for $\theta > 0$) or the left side (for $\theta < 0$) of the chain. Such property is reminiscent of mechanical topological insulators and has been reported in linkages and origami structures. In the context of the present study, where we actuate the chain either from the left or the right side, this property implies that the chain either amplifies or absorbs the input displacement.

If we assume that the spatial variations are small between neighboring unit cells, the discrete set of Equations (39) can be considered in the continuum limit. Therefore we define $b \to dx$, $u_n \to u(x)$ and taking a Taylor expansion around $x = n$, $u_{n+1} \to u(x) + b(du/dx)$, we obtain

$$0 = -\sin(2\theta)\, u + \sqrt{2}a\cos\theta\, (2 + \cos(2\theta) - \sin(2\theta))\, \frac{du}{dx}, \tag{42}$$

### 2. Nonlinear behavior

In the main text, we investigate the nonlinear response of the mechanical metamaterial. To this end, we solve the systems of Equations (34-36) numerically and obtain $u_n(u_0)$.

### C. Maxwell-Betti reciprocity theorem

In order to probe deviations from Maxwell-Betti reciprocity theorem, we assume that that only vertices $U$ have torsional springs of torsional stiffness $C$ (depicted by the red dots in Fig. S3). We have verified that this simplifying assumption does not affect the following reasoning. This assumption allows us to simply express the elastic energy stored in the structure as

$$E = \frac{C}{2} \sum_{n=0}^{N} 4\delta\theta_n^2, \tag{43}$$

where $\delta\theta_n$ are the solutions of Eqs. (34-36). On the one hand, when the chain is actuated under boundary condition $(A)$ from the left on node $n = 0$ with an input displacement $u_0$, the reaction force is given by

$$F_{0(A)} \equiv \frac{\partial E}{\partial u_0} = 4C \sum_{n=0}^{N} \frac{\partial \delta\theta_n}{\partial u_0} \delta\theta_n. \tag{44}$$

On the other hand, when the chain is actuated from the right under boundary condition $(B)$ on node $N$ with an input displacement $u_N$, the reaction force is given by

$$F_{N(B)} \equiv \frac{\partial E}{\partial u_N} = 4C \sum_{n=0}^{N} \frac{\partial \delta\theta_n}{\partial u_N} \delta\theta_n. \tag{45}$$

It is straightforward to demonstrate that the solution to the linearized equation satisfies Maxwell-Betti theorem. After a few simple manipulations, we obtain $u_{A \to B} \equiv u_{N(A)} = F_{0(A)} K(\theta)$ and $u_{B \to A} \equiv u_{0(B)} = F_{N(B)} K(\theta)$, with

$$K(\theta) = \frac{2a^2 \cos^2\theta}{C \sum_{n=0}^{N} g(\theta)^{2n-N}}. \tag{46}$$

As $u_{B \to A} F_{0(A)} = u_{A \to B} F_{N(B)} = K(\theta) F_{0(A)} F_{N(B)}$, Maxwell-Betti theorem is verified and the structure is reciprocal.

To probe the non-reciprocal behavior and compute $\Delta u = u_{A \to B} + u_{B \to A}$ within the nonlinear response, we combine our numerical solution for the displacement field $u_n(u_0)$ with Eq. (43), which we differentiate numerically. We have checked that the numerical and linearized solutions coincide in the limit of small input forces. The results are displayed in Figs. 2e-h of the main text.

### D. Topological invariant : winding number

In this last section, we calculate the topological invariant characterizing the response of our metamaterial. As mentioned above, our metamaterial of finite length has only one soft mode of deformation, as it has only one degree of freedom ($3N + 2$) more than constraints ($3N + 1$). However, a periodic version of such metamaterial would have exactly the same number of degrees of freedom ($3N + 3$) as constraints ($3N + 3$). Such a system is called isostatic and its topological invariant—called the winding number—can be computed directly using the framework developed by Kane & Lubensky[2] for isostatic lattices. This framework consists in mapping the mechanical problem onto a quantum-mechanical problem, for which the topological properties are well established. Such mapping is greatly simplified in the case of isostatic mechanical systems, and the winding number can be readily calculated by considering the so-called compatibility matrix, which relates the internal motions to the internal changes of bond length in the system.

To calculate the compatibility matrix for the system under scrutiny and calculate the topological properties of the metamaterial, we assume that the link connecting the unit cell $n$ to the unit cell $n + 1$ is extensible of stiffness $k$, in contrast with our analysis above where it was assumed infinitely rigid. In the limit where this stiffness $k$ is much larger than the torsional stiffness $C$, the following calculations remain fully consistent with the above analysis.

We denote the relative length change of such link $\varepsilon_{n,n+1}$. Therefore the vector describing such link is

$$\ell'_{\mathbf{n,n+1}} = (1 + \varepsilon_{n,n+1})\ell_{\mathbf{n,n+1}}, \tag{47}$$

where $\ell_{\mathbf{n,n+1}}$ is defined in eq. (32). This new hypothesis modifies Eqs. (34-36), which can be linearized and rewritten to express the relative length change of the link $\varepsilon_{n,n+1}$ as a function of the angles $\delta\theta_n$ and $\delta\theta_{n+1}$. We carry out the aforementioned steps and find

$$\varepsilon_{n,n+1} = c_1 \delta\theta_n - c_2 \delta\theta_{n+1}, \tag{48}$$

where $c_1 = (\cos(2\theta) + \sin(2\theta) + 2)/2$ and $c_2 = (\cos(2\theta) - \sin(2\theta) + 2)/2$. This equation is a sub-block of the compatibility matrix, which for the entire system relates the links relative length changes $\varepsilon_{n,n+1}$ to the angles $\delta\theta_n$ as follows

$$\underbrace{\begin{bmatrix} \varepsilon_{0,1} \\ \varepsilon_{1,2} \\ \vdots \\ \varepsilon_{n,n+1} \\ \vdots \\ \varepsilon_{N-1,N} \end{bmatrix}}_{1 \times N} = \underbrace{\begin{bmatrix} \delta\theta_0 \\ \delta\theta_1 \\ \vdots \\ \vdots \\ \delta\theta_n \\ \vdots \\ \delta\theta_N \end{bmatrix}}_{1 \times (N+1)} \underbrace{\begin{bmatrix} c_1 & -c_2 & 0 & \cdots & \cdots & 0 & 0 \\ 0 & c_1 & -c_2 & 0 & \cdots & 0 & 0 \\ 0 & \ddots & \ddots & \ddots & \ddots & \vdots & \vdots \\ \vdots & & \ddots & c_1 & -c_2 & 0 & 0 \\ \vdots & & & \ddots & c_1 & -c_2 & 0 \\ 0 & \cdots & \cdots & \cdots & 0 & c_1 & -c_2 \end{bmatrix}}_{(N+1) \times N}. \tag{49}$$

The $(N+1) \times N$ above is called the compatibility matrix $\mathbf{C}$. In order to calculate the winding number from this equation, we assume that the system is infinite and perform a Fourier analysis. In an infinite system, the links relative length changes $\varepsilon_{n,n+1}$ and the rotation angles $\delta\theta_n$ can be expressed as plane waves

$$\varepsilon_{n,n+1} = \hat{\varepsilon}(q)e^{iqn}, \tag{50}$$

and

$$\delta\theta_n = \hat{\delta\theta}(q)e^{iqn}. \tag{51}$$

As a result of the translation properties of the Fourier transform, equation 48 becomes, for each value of $q$,

$$\hat{\varepsilon}(q) = C(q)\hat{\delta\theta}(q), \tag{52}$$

where

$$C(q) = c_1 - c_2 e^{iq} \tag{53}$$

can be interpreted as the Fourier transform of the compatibility matrix. $C(q)$ completes a circle in the complex plane as the wave vector $q$ cycles along the Brillouin zone, namely between 0 and $2\pi$. The winding number $W$ is equal to 0 or 1 depending on whether the circle encircles the origin. Using the graphical representation shown in the insets of

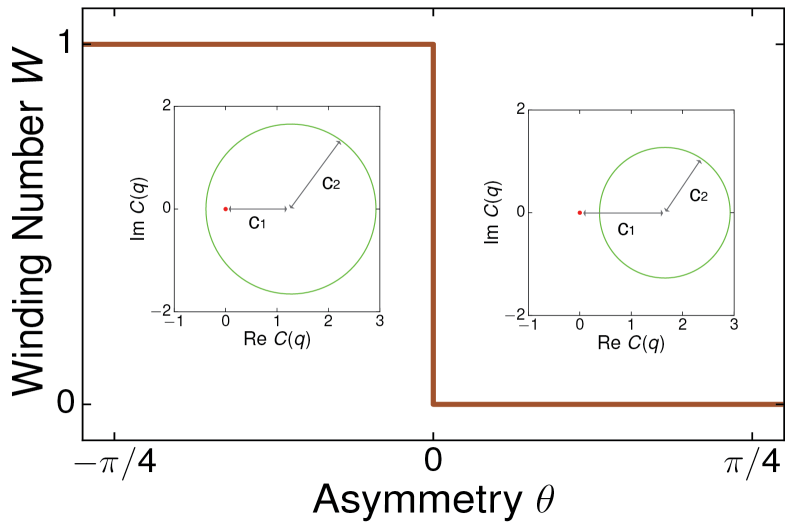

FIGURE S4. Topological properties of the mechanical metamaterial 2. Winding number $W$ vs. asymmetry angle $\theta$. Insets : parametric plots of $C(q)$ in the complex plane as the wave vector $q$ cycles along the Brillouin zone for an asymmetry angle $\theta = -\pi/16$ (left) and $\theta = \pi/16$ (right). In the left inset, $\theta < 0$ (equivalently, $c_2 > c_1$), the contour $C(q)$ does not encircle the origin (denoted by a red dot) and $W = 1$. In the right inset, $\theta > 0$ (equivalently, $c_2 < c_1$), the contour $C(q)$ encircles the origin (denoted by a red dot) and $W = 0$.

Fig. S4, it follows directly from Eq. (53) that $W = 1$ when $c_2 > c_1$ (or equivalently when $g(\theta) = c_1/c_2 < 1$) and $W = 0$ when $c_2 < c_1$ (or equivalently when $g(\theta) > 1$). An equivalent yet more formal way to define the winding number is[1,2]

$$W = \frac{1}{2\pi i} \oint_0^{2\pi} dq \frac{C'(q)}{C(q)}. \tag{54}$$

Both approaches lead to the same result, namely

$$W = \begin{cases} 1 & \text{if } \theta < 0 \\ 0 & \text{if } \theta > 0 \end{cases} \tag{55}$$

Therefore the value of the winding number is utterly related to the localized edge mode described above, equal to 1 (respectively 0) when the soft mode is localized on the left (respectively right) side of the chain. This confirms the topological nature of the linear response of the metamaterial. We summarize these findings in Fig. S4.

### E. Topological edge mode

In this part, we show that the topological nature of the response manifests itself in the form of a unique edge mode, which is localised either on the left or right side of the metamaterial, depending on the asymmetry angle $\theta$. To do this, we calculate the dynamical matrix associated to the compatbility matrix $\mathbf{D}$ as follows[1]

$$\mathbf{D} = \frac{k}{\Gamma}\mathbf{C}^T\mathbf{C} + \frac{C}{\Gamma}\mathbb{1}_{(N+1)\times(N+1)} \tag{56}$$

where $k$ is the stiffness of the links, $\Gamma$ is the moment of inertia of the rotating squares, $C$ is the torsional stiffness and $\mathbb{1}_{(N+1)\times(N+1)}$ is the identity matrix. In the limit where $k \gg C$, which is relevant for the description of our metamaterial, the second term of this equation can be neglected we obtain thus

$$\mathbf{D} = \frac{k}{\Gamma} \underbrace{\begin{bmatrix} c_1^2 & -c_1 c_2 & 0 & \cdots & \cdots & 0 \\ -c_1 c_2 & c_1^2 + c_2^2 & -c_1 c_2 & 0 & \cdots & 0 \\ 0 & \ddots & \ddots & \ddots & \ddots & \vdots \\ \vdots & & \ddots & c_1^2 + c_2^2 & -c_1 c_2 & 0 \\ \vdots & & & \ddots & c_1^2 + c_2^2 & -c_1 c_2 \\ 0 & \cdots & \cdots & \cdots & -c_1 c_2 & c_2^2 \end{bmatrix}}_{(N+1)\times(N+1)}. \tag{57}$$

The vibrational modes of the metamaterial can be directly analysed from the spectral analysis of the dynamical matrix, where the eigenvalues are the square of the eigen-frequencies. As mentioned above, the most straightforward method to investigate such vibrational properties is to assume at first the system infinite and perform a Fourier analysis. In an infinite system, the field $\delta\theta_n$ can be expressed as plane waves

$$\delta\theta_n = \hat{\delta\theta}(q)e^{iqn}, \tag{58}$$

As a result of the translation property of the Fourier transform, we obtain

$$0 = \hat{\delta\theta}(q)D(q), \tag{59}$$

where

$$D(q) = \omega_0^2 \left(c_1^2 + c_2^2 - 2c_1 c_2 \cos q\right) \tag{60}$$

is a scalar that can be interpreted as the Fourier transform of the dynamical matrix. $\omega_0 = \sqrt{k/\Gamma}$ is the characteristic frequency of the structure. We plot in Fig. S5a the corresponding band diagram, which, noticeably, exhibits a band-gap $\Delta\omega$. The smoking gun of topologically non-trivial response in static structures is the presence of a vanishing band-gap in the phonon spectrum centered at zero-wave vector, which is often associated to a change of the topological properties[2]. In accordance with the topological transition reported above, we see again here that the dynamical matrix $D(q=0)$ becomes zero when the coefficients $c_1$ and $c_2$ are equal.

To show explicitly that it corresponds to zero-energy edge modes, we perform the spectral analysis for a finite system size and we summarize our findings in Fig. S5b-d. This analysis reveals that the finite system has one edge modes of zero-frequency (solid red line in Fig. S5b and red bar in S5c), which by contrast with the rest of the spectrum (gray lines in Fig. S5b and gray bars in Fig. S5c) is comprised in the band-gap. For the present value of the structure asymmetry $\theta = \pi/16$ and for all positive values of $\theta$ in general, the edge mode is localised on the right. We find that it is localised on the left for negative values of $\theta$. We calculate the frequency of such edge mode for a continuous range of asymmetry angles $\theta$ and find that it is always zero (See Fig. S5d). The angle $\theta = 0$ is peculiar as it corresponds to the point where the band-gap closes and where the topological transition occurs.

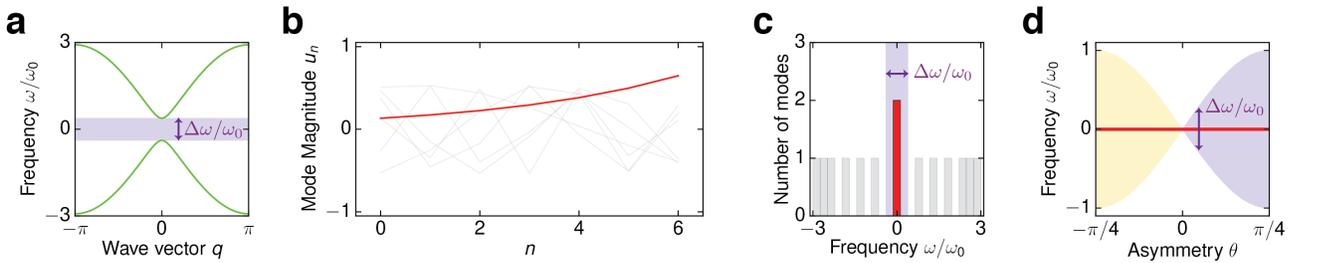

FIGURE S5. Mode analysis of the (topological) mechanical metamaterial II. (a) Rescaled frequency $\omega/\omega_0$ vs. wave vector $q$ from the Fourier analysis carried out by assuming that the structure infinite. The purple shaded area denotes the band-gap. (b) Spatial profile of the eigenmodes of the finite structure ($N = 6$). The eigenmode with the lowest frequency is displayed in solid red and corresponds to the edge-mode. The higher frequency eigenmodes are displayed in solid light gray lines. (c) Histogram of the rescaled eigen-frequencies. The edge-mode located in the band-gap (purple shaded area) is shown in red and the remaining "bulk" modes in light gray. $\omega_0 = \sqrt{k/\Gamma}$ is the characteristic frequency of the structure. (d) Band-gap (purple and yellow shaded areas) and edge modes rescaled frequencies $\omega_E/\omega_0$ (thick red line) vs. asymmetry angle $\theta$ of the structure. All panels have been produced using $N = 6$, and $a = 1$, and panels (a-c) have been produced using $\theta = \pi/16$.

———————————————

[1] T.C. Lubensky, C.L. Kane, X. Mao, A. Souslov and K. Sun, *Phonons and elasticity in critically coordinated lattices*, Rep. Prog. Phys. **78**, 073901 (2015).
[2] C.L. Kane and T.C. Lubensky, *Topological boundary modes in isostatic lattices*, Nat. Phys. **10**, 39-45 (2014).
[3] M. Z. Hasan and C.L. Kane, *Colloquium : Topological insulators*, Rev. Mod. Phys. **82**, 3045 (2010).